\documentclass[runningheads]{llncs}
\usepackage{graphicx}
\usepackage{booktabs}
\usepackage{array}
\usepackage{amsmath}
\usepackage{caption}
\usepackage{subcaption}
\usepackage{diagbox}
\usepackage{amssymb}
\usepackage{floatrow}
\usepackage[table,xcdraw]{xcolor}
\usepackage{blindtext}
\usepackage[export]{adjustbox}
\usepackage{multirow}
\usepackage{hyperref}
\usepackage{makecell}

\newfloatcommand{capbtabbox}{table}[][\FBwidth]

\newcommand{\set}[1]{\mathcal{#1}}
\newcolumntype{L}[1]{>{\PreserveBackslash\raggedright}p{#1}}

\newcommand{\numberutterances}{$\#_{turns}$}
\newcommand{\avgnumwordsq}{$\overline{\#_{\set{U}words}}$}
\newcommand{\avgnumwordsd}{$\overline{\#_{\set{R}words}}$}
\newcommand{\stdsm}{$\sigma_{SM}$}
\newcommand{\stdbm}{$\sigma_{BM25}$}
\newcommand{\avgbertpred}{$BERT_{pred}$}
\newcommand{\avgbertloss}{$\overline{BERT_{loss}}$}

\newcommand{\mantis}{\texttt{MANtIS}}
\newcommand{\msdialog}{\texttt{MSDialog}}
\newcommand{\stack}{Stack Exchange}

\newcommand{\pacing}{f_{pace}}
\newcommand{\scoring}{f_{score}}

\title{Curriculum Learning Strategies for IR}
\subtitle{An Empirical Study on Conversation Response Ranking}

\author{Gustavo Penha \and
Claudia Hauff}
\institute{TU Delft\\
\email{\{g.penha-1,c.hauff\}@tudelft.nl}}

\begin{document}

\maketitle              % typeset the header of the contribution
\begin{abstract}
Neural ranking models are traditionally trained on a series of random batches, sampled uniformly from the entire training set. Curriculum learning has recently been shown to improve neural models' effectiveness by sampling batches non-uniformly, going from easy to difficult instances during training. In the context of neural Information Retrieval~(IR) curriculum learning has not been explored yet, and so it remains unclear (1) how to \emph{measure the difficulty} of training instances and (2) \emph{how to transition} from easy to difficult instances during training. To address both challenges and determine whether curriculum learning is beneficial for neural ranking models, we need large-scale datasets and a retrieval task that allows us to conduct a wide range of experiments. For this purpose, we resort to the task of \emph{conversation response ranking}: ranking responses given the conversation history. In order to deal with challenge (1), we explore \emph{scoring functions} to measure the difficulty of conversations based on different input spaces. To address challenge (2) we evaluate different \emph{pacing functions}, which determine the velocity in which we go from easy to difficult instances. We find that, overall, by just intelligently sorting the training data (i.e., by performing curriculum learning) we can improve the retrieval effectiveness by up to 2\%\footnote{The source code is available at~\url{https://github.com/Guzpenha/transformers\_cl}.}. 
\end{abstract}

\section{Introduction}

Curriculum Learning (CL) is motivated by the way humans teach complex concepts: teachers impose a certain order of the material during students' education. Following this guidance, students can exploit previously learned concepts to more easily learn new ones. This idea was initially applied to machine learning over two decades ago~\cite{elman1993learning} as an attempt to use a similar strategy in the training of a recurrent network by \textit{starting small} and gradually learning more difficult examples. More recently, Bengio et al.~\cite{bengio2009curriculum} provided additional evidence that curriculum strategies can benefit neural network training with experimental results on different tasks such as shape recognition and language modelling. Since then, empirical successes were observed for several computer vision~\cite{hacohen2019power,weinshall2018curriculum} and natural language processing (NLP) tasks~\cite{sachan2016easy,rajeswar2017adversarial,zhang2018}.

In supervised machine learning, a function is learnt by the learning algorithm (the \textit{student}) based on inputs and labels provided by the \textit{teacher}. The teacher typically samples randomly from the entire training set. In contrast, CL imposes a structure on the training set based on a notion of difficulty of instances, presenting to the student easy instances before difficult ones. When defining a CL strategy we face two challenges that are specific to the domain and task at hand~\cite{hacohen2019power}: (1) arranging the training instances by a sensible measure of \emph{difficulty}, and, (2) determining the \emph{pace} in which to present instances---going over easy instances too fast or too slow might lead to ineffective learning.

We conduct here an empirical investigation into those two challenges in the context of IR. Estimating relevance---a notion based on human cognitive processes---is a complex and difficult task at the core of IR, and it is still unknown \emph{to what extent CL strategies are beneficial for neural ranking models}. This is the question we aim to answer in our work.

Given a set of queries---for instance user utterances, search queries or questions in natural language---and a set of documents---for instance responses, web documents or passages---neural ranking models learn to distinguish relevant from non-relevant query-document pairs by training on a large number of labeled training pairs. Neural models have for some time struggled to display significant and additive gains in IR~\cite{Yang:2019:CEH:3331184.3331340}. In a short time though, BERT~\cite{devlin2019bert} (released in late 2018) and its derivatives (e.g. XLNet~\cite{yang2019xlnet}, RoBERTa~\cite{liu2019roberta}) have proven to be remarkably effective for a range of NLP tasks. The recent breakthroughs of these large and heavily pre-trained language models have also benefited IR~\cite{yang2019simple,yang2019end,yilmaz2019cross}.

In our work we focus on the challenging IR task of conversation response ranking~\cite{wu2017sequential}, where the query is the dialogue history and the documents are the candidate responses of the agent. The set of responses are not generated on the go, they must be retrieved from a comprehensive dialogue corpus. A number of deep neural ranking models have recently been proposed for this task~\cite{tao2019one,yang2018response,zhang2018modeling,wu2017sequential,zhou2018multi}, which is more complex than retrieval for single-turn interactions, as the ranking model has to determine where the important information is in the previous user utterances (dialogue history) and how it is relevant to the current information need of the user. Due to the complexity of the relevance estimation problem displayed in this task, we argue it to be a good test case for curriculum learning in IR.

In order to tackle the first challenge of CL (determine what makes an instance difficult) we study different \emph{scoring functions} that determine the difficulty of query-document pairs based on four different input spaces: conversation history \{$\set{U}$\}, candidate responses $\{\set{R}\}$, both $\{\set{U}$,$\set{R}\}$, and $\{\set{U}$, $\set{R}$, $\set{Y}\}$, where $\set{Y}$ are relevance labels for the responses. To address the second challenge (determine the pace to move from easy to difficult instances) we explore different \emph{pacing functions} that serve easy instances to the learner for more or less time during the training procedure. We empirically explore how the curriculum strategies perform for two different response ranking datasets when compared against vanilla (no curriculum) fine-tuning of BERT for the task. Our main findings are that (i) CL improves retrieval effectiveness when we use a difficulty criteria based on a supervised model that uses all the available information $\{\set{U}$, $\set{R}$, $\set{Y}\}$, (ii) it is best to give the model more time to assimilate harder instances during training by introducing difficult instances in earlier iterations, and, (iii) the CL gains over the no curriculum baseline are spread over different conversation domains, lengths of conversations and measures of conversation difficulty.
\section{Related Work}

\subsubsection*{Neural Ranking Models}
Over the past few years, the IR community has seen a great uptake of the many flavours of deep learning for all kinds of IR tasks such as ad-hoc retrieval, question answering and conversation response ranking. Unlike traditional learning to rank (LTR)~\cite{liu2009learning} approaches in which we manually define features for queries, documents and their interaction, neural ranking models learn features directly from the raw textual data. Neural ranking approaches can be roughly categorized into representation-focused~\cite{huang2013learning,shen2014latent,wan2016deep} and interaction-focused~\cite{guo2016deep,wan2016match}. The former learns query and document representations separately and then computes the similarity between the representations. In the latter approach, first a query-document interaction matrix is built, which is then fed to neural net layers. Estimating relevance directly based on interactions, i.e. interaction-focused models, has shown to outperform representation-based approaches on several tasks~\cite{nie2018empirical,hu2014convolutional}. 

Transfer learning via large pre-trained Transformers~\cite{vaswani2017attention}---the prominent case being BERT~\cite{devlin2019bert}---has lead to remarkable empirical successes on a range of NLP problems. The BERT approach to learn textual representations has also significantly improved the performance of neural models for several IR tasks~\cite{yang2019simple,yang2019end,sakata2019faq,Qu:2019:BHA:3331184.3331341,yilmaz2019cross}, that for a long time struggled to outperform classic IR models~\cite{Yang:2019:CEH:3331184.3331340}. In this work we use the no-CL BERT as a strong baseline for the conversation response ranking task.
% Transformer-based architectures for IR, that concatenate the query and the document and use it as input, can be categorized as interaction-focused neural ranking models since the self-attention mechanism of BERT learns interaction between all words in the input string. 

\subsubsection*{Curriculum Learning}

Following a curriculum that dictates the ordering and content of the education material is prevalent in the context of human learning. With such guidance, students can exploit previously learned concepts to ease the learning of new and more complex ones. Inspired by cognitive science research~\cite{rohde1999language}, researchers posed the question of whether a machine learning algorithm could benefit, in terms of learning speed and effectiveness, from a similar curriculum strategy \cite{elman1993learning,bengio2009curriculum}. Since then, positive evidence for the benefits of curriculum training, i.e. training the model using easy instances first and increasing the difficulty during the training procedure, has been empirically demonstrated in different machine learning problems, e.g. image classification~\cite{hacohen2019power,gong2016multi}, machine translation~\cite{platanios2019competence,kocmi2017curriculum,zhang2018} and answer generation~\cite{liu2018curriculum}. 

Processing training instances in a meaningful order is not unique to CL. Another related branch of research focuses on \emph{dynamic} sampling strategies~\cite{kumar2010self,chang2017active,shrivastava2016training,breiman1998arcing}, which unlike CL that requires a definition of what is easy and difficult before training starts, estimates the importance of instances during the training procedure. Self-paced learning~\cite{kumar2010self} simultaneously selects easy instances to focus on and updates the model parameters by solving a biconvex optimization problem. A seemingly contradictory set of approaches give more focus to difficult or more uncertain instances. In active learning~\cite{cohn1996active,tong2001support,chang2017active}, the most uncertain instances with respect to the current classifier are employed for training. Similarly, hard example mining~\cite{shrivastava2016training} focuses on difficult instances, measured by the model loss or magnitude of gradients for instance. Boosting~\cite{breiman1998arcing,zhang2017boosting} techniques give more weight to difficult instances as training progresses. In this work we focus on CL, which has been more successful in neural models, and leave the study of dynamic sampling strategies in neural IR as future work.

The most critical part of using a CL strategy is defining the difficulty metric to sort instances by. The estimation of instance difficulty is often based on our prior knowledge on what makes each instance difficult for a certain task and thus is domain dependent (cf. Table~\ref{table:related_work_dif_measures} for curriculum examples). CL strategies have not been studied yet in neural ranking models. To our knowledge, CL has only recently been employed in IR within the LTR framework, using LambdaMart~\cite{burges2010ranknet}, for ad-hoc retrieval by Ferro et al.~\cite{ferro2018continuation}. However, no effectiveness improvements over randomly sampling training data were observed. The representation of the query, document and their interactions in the traditional LTR framework is dictated by the manually engineered input features. We argue that neural ranking models, which learn how to represent the input, are better suited for applying CL in order to learn increasingly more complex concepts. \vspace{-0.5cm}

\begin{table}[!htb]
\small
\caption{Difficulty measures used in the curriculum learning literature.}
\centering
\label{table:related_work_dif_measures}
\begin{tabular}{@{}lp{8.2cm}@{}}
\toprule
\multicolumn{1}{l}{\textbf{Difficulty criteria}} & \multicolumn{1}{l}{\textbf{Tasks}} \\ \midrule
sentence length & machine translation~\cite{platanios2019competence},  language generation ~\cite{rajeswar2017adversarial}, reading comprehension ~\cite{yu2016end} \\ \midrule
word rarity & machine translation~\cite{platanios2019competence,zhang2018}, language modeling~\cite{bengio2009curriculum} \\ \midrule
external model confidence & machine translation~\cite{zhang2018}, image classification~\cite{weinshall2018curriculum,hacohen2019power},  ad-hoc retrieval~\cite{ferro2018continuation}\\ \midrule
supervision signal intensity & facial expression recognition~\cite{gui2017curriculum}, ad-hoc retrieval~\cite{ferro2018continuation}\\  \midrule
noise estimate & speaker identification~\cite{ranjan2018curriculum}, image classification~\cite{chen2015webly} \\ \midrule
human annotation & image classification~\cite{tudor2016hard} (through weak supervision) \\
\bottomrule
\end{tabular}
\vspace{-1cm}
\end{table}
\section{Curriculum Learning}

Before introducing our experimental framework (i.e., the scoring functions and the pacing functions we investigate), let us first formally introduce the specific IR task we explore---a choice dictated by the complex nature of the task (compared to e.g. ad-hoc retrieval) as well as the availability of large-scale training resources such as \msdialog{}~\cite{qu2018analyzing} and UDC~\cite{lowe2015ubuntu}.

\subsubsection{Conversation Response Ranking}

Given a historical dialogue corpus and a conversation, (i.e., the user's current utterance and the conversation history) the task of conversation response ranking~\cite{wu2017sequential,yang2018response,tao2019one} is defined as the ranking of the most relevant response available in the corpus. This setup relies on the fact that a large corpus of historical conversation data exists and adequate replies (that are coherent, well-formulated and informative) to user utterances can be found in it~\cite{yang2019hybrid}. Formally, let $\set{D}=\{(\set{U}_i, \set{R}_i, \set{Y}_i)\}_{i=1}^{N}$ be an information-seeking conversations data set consisting of $N$ triplets: dialogue context, response candidates and response labels. The dialogue context $\set{U}_i$ is composed of the previous utterances $\{u^1, u^2, ... , u^{\tau}\}$ at the turn $\tau$ of the dialogue.  The candidate responses $\set{R}_i = \{r^1, r^2, ..., r^k\}$ are either the true response ($u^{\tau+1}$) or negative sampled candidates\footnote{In a production setup the ranker would either retrieve responses from the entire corpus or re-rank the responses retrieved by a recall-oriented retrieval method.}. The relevance labels $\set{Y}_i = \{y^1, y^2, ..., y^k\}$ indicate the responses' binary relevance scores, 1 if $r = u^{\tau+1}$ and 0 otherwise. The task is then to learn a ranking function $f(.)$ that is able to generate a ranked list for the set of candidate responses $\set{R}_i$ based on their predicted relevance scores $f(\set{U}_i,r)$.

\begin{figure}[!htb]
    \centering
    \includegraphics[width=1.\textwidth]{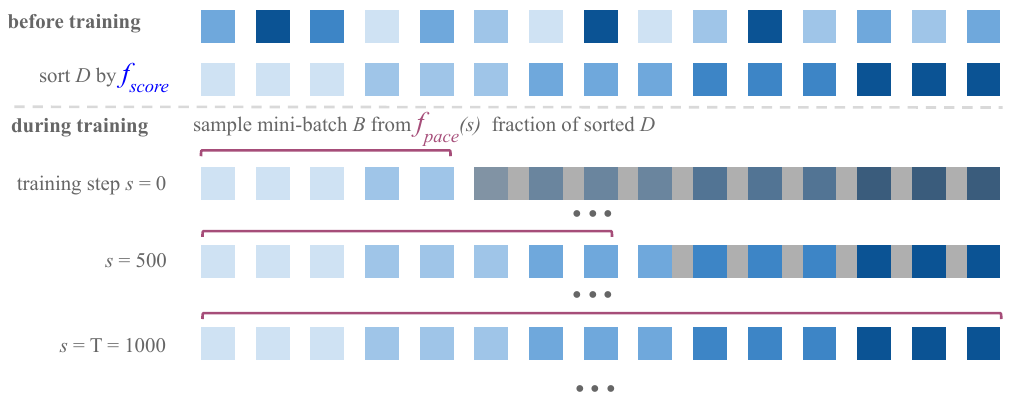}
    \caption{Our curriculum learning framework is defined by two functions. The scoring function $\scoring{}(instance)$ defines the instances' difficulty (darker/lighter blue indicate higher/lower difficulty). The pacing function $\pacing{}(s)$ indicates the percentage of the dataset available for sampling according to the training step $s$.}
    \label{fig:framework}
\end{figure}

\vspace{-1cm}
\subsubsection{Curriculum Framework}
When training neural networks, the common training procedure is to divide the dataset $\set{D}$ into $\set{D}_{train}, \set{D}_{dev}, \set{D}_{test}$ and randomly (i.e., uniformly---every sample has the same likelihood of being sampled) sample mini-batches $\set{B}=\{(\set{U}_i, \set{R}_i, \set{Y}_i)\}_{i=1}^{k}$ of $k$ instances from $\set{D}_{train}$ where $k\ll N$, and perform an optimization procedure sequentially in $\{\set{B}_1,...,\set{B}_M\}$. The CL framework employed here is inspired by previous works~\cite{weinshall2018curriculum,platanios2019competence}. It is defined by two functions: the \emph{scoring function} which determines the difficulty of instances and the \emph{pacing function} which controls the pace with which to transition from easy to hard instances during training. More specifically, the scoring function $\scoring{}(\set{U}_i, \set{R}_i, \set{Y}_i)$, is used to sort the training dataset. The pacing function $\pacing{}(s)$ determines the percentage of the sorted dataset available for sampling according to the current training step $s$ (one forward pass plus one backward pass of a batch is considered to be one step). The neural ranking model samples uniformly from the initial $\pacing{}(s) * |{D}_{train}|$ instances sorted by $\scoring{}$, while the rest of the dataset is not available for sampling. During training $\pacing{}(s)$ goes from $\delta$ (percentage of initial training data) to 1 when $s=T$. Both $\delta$ and $T$ are hyperparameters. We provide an illustration of the training process in Figure \ref{fig:framework}. 

\subsubsection{Scoring Functions}

\begin{table}[]
\small
\centering
\caption{Overview of our curriculum learning scoring functions. 
}
\label{table:scoring_functions_definition}
\begin{tabular}{p{1.5cm}llp{2cm}}
\toprule \textbf{Input Space} & \textbf{Name} & \textbf{Definition} & \textbf{Difficulty notion} \\ \midrule
 baseline & \textit{random} & $\scoring{} = Uniform(0,1)$ & \\ \midrule
\multirow{2}{*}{$(\set{U})$} & \numberutterances  & $\scoring{}(\set{U}) = |\set{U}|$ & \multirow{2}{2cm}{information spread}\\ 
 &\avgnumwordsq  & $\scoring{}(\set{U}) = \frac{\sum_{i=0}^{|\set{U}|} word\_count(u_i)}{|\set{U}|}$ \\ \midrule
$(\set{R})$ & \avgnumwordsd  & $\scoring{}(\set{R}) = \frac{\sum_{i=0}^{|\set{R}|} word\_count(r_i)}{|\set{R}|}$ & distraction in responses \\ \midrule
\multirow{2}{*}{$(\set{U},\set{R})$} & \stdsm & $\scoring{}(\set{U},\set{R}) = \sqrt{\frac{\sum_{i=0}^{|\set{R}|} (SM(\set{U},r_{i})-\overline{SM(\set{U},\set{R})})^2}{|\set{R}|-1}}$ & \multirow{2}{2cm}{responses heterogeneity} \\
&\stdbm & $\scoring{}(\set{U},\set{R}) = \sqrt{\frac{\sum_{i=0}^{|\set{R}|} (BM25(\set{U},r_{i})-\overline{BM25(\set{U},\set{R})})^2}{|\set{R}|-1}}$ \\ \midrule
\multirow{2}{*}{$(\set{U},\set{R},\set{Y})$}&\avgbertpred & 
$\!\begin{aligned}[t]
&\scoring{}(\set{U},\set{R},\set{Y}) = \\
&- (BERT\_pred(\set{U},r_{i}^{+}) -BERT\_pred(\set{U},r_{i}^{-}))
\end{aligned}$
& \multirow{4}{*}{\makecell{model \\ confidence}} \\
&\avgbertloss & $\scoring{}(\set{U},\set{R},\set{Y}) = \frac{\sum_{i=0}^{|\set{R}|} BERT\_loss(\set{U},r_{i})}{|\set{R}|}$ \\ \bottomrule
\end{tabular}
\vspace{-2mm}%
\end{table}

In order to measure the difficulty of a training triplet composed of $(\set{U}_i, \set{R}_i, \set{Y}_i)$, we define pacing functions that use different parts of the input space: functions that leverage (i) the text in the dialogue history $\{\set{U}\}$ (ii) the text in the response candidates $\{\set{R}\}$ (iii) interactions between them, i.e., $\{\set{U},\set{R}\}$, and, (iv) all available information including the labels for the training set, i.e., $\{\set{U},\set{R},\set{Y}\}$. The seven\footnote{The function \textit{random} is the baseline---instances are sampled uniformly (no CL).} scoring functions we propose are defined in Table \ref{table:scoring_functions_definition}; we now provide intuitions of why we believe each function to capture some notion of instance difficulty.

\begin{itemize}
\setlength\itemsep{1em} 

\item[$\bullet$] \numberutterances $(\set{U})$ and \avgnumwordsq $(\set{U})$: The important information in the context can be spread over different utterances and words. Bigger dialogue contexts means there are more places where the important part of the user information need can be spread over. \avgnumwordsd $(\set{R})$: Longer responses can distract the model as to which set of words or sentences are more important for matching. Previous work shows that it is possible to fool machine reading models by creating longer documents with additional distracting sentences~\cite{jia2017adversarial}.

\item[$\bullet$] \stdsm $(\set{U,R})$ and \stdbm $(\set{U,R})$: Inspired by query performance prediction literature~\cite{shtok2009predicting}, we use the variance of retrieval scores to estimate the amount of heterogeneity of information, i.e. diversity, in the response candidate. Homogeneous ranked lists are considered to be easy. We deploy a semantic matching model (SM) and BM25 to capture both semantic correspondences and keyword matching~\cite{jinfeng2019bridging}. SM is the average cosine similarity between the first $k$ words from $\set{U}$ (concatenated utterances) with the first $k$ words from $r$ using pre-trained word embeddings. 

\item[$\bullet$] \avgbertpred $(\set{U,R,Y})$ and \avgbertloss $(\set{U,R,Y})$ : Inspired by CL literature~\cite{hacohen2019power}, we use external model prediction confidence scores as a measure of difficulty\footnote{We note, that using BM25 average precision as a scoring function failed to outperform the baseline.}. We fine-tune BERT~\cite{devlin2019bert} on $\set{D}_{train}$ for the conversation response ranking task. For \avgbertpred{} easy dialogue contexts are the ones that the BERT confidence score for the positive response $r^{+}$ candidate is higher than the confidence for the negative response candidate $r^{-}$. The higher the difference the easier the instance is. For \avgbertloss we consider the loss of the model to be an indicator of the difficulty of an instance.

\end{itemize}

\subsubsection*{Pacing functions}

\begin{figure}
% \captionsetup{margin=1cm}
\begin{floatrow}
\capbtabbox{%
\centering
\scriptsize
\tiny
\begin{tabular}{@{}ll@{}}
\toprule
\multicolumn{1}{c}{\textbf{Pacing function}} & \multicolumn{1}{c}{\textbf{Definition}} \\ \midrule
\textit{baseline\_training} & $\pacing{}(s) = 1$ \\ \midrule
\textit{step} &
    $\pacing{}(s) =
     \begin{cases}
       \delta, & \text{if}\ s \le T*0.33 \\
       0.66, & \text{if}\ s>T*0.33, s \le T*0.66\\
       1, & \text{if}\ s > T*0.66\\
     \end{cases} $\\ \midrule
\textit{root} & $\pacing{}(s,n) = min \left(1,\left(s \frac{1-\delta^{n}}{T}+\delta^{n}\right)^{\frac{1}{n}}\right)$ \\ 
\textit{linear} & $\pacing{}(s,n) = root(s,1)$\\
\textit{root\_n} & $\pacing{}(s,n) = root(s,n)$ \\ \midrule
\textit{geom\_progression} & $\pacing{}(s) =min \left(1,2^{\left(s \frac{log_21-log_2\delta}{T}+log_2\delta\right)}\right)$\\  
\bottomrule
\end{tabular}
}{%
  \caption{Overview of our curriculum learning pacing functions. $\delta$ and $T$ are hyperparameters.}
\label{table:pacing_functions_definition}%
}
\ffigbox[0.35\textwidth]{%
  \includegraphics[width=0.35\textwidth,left]{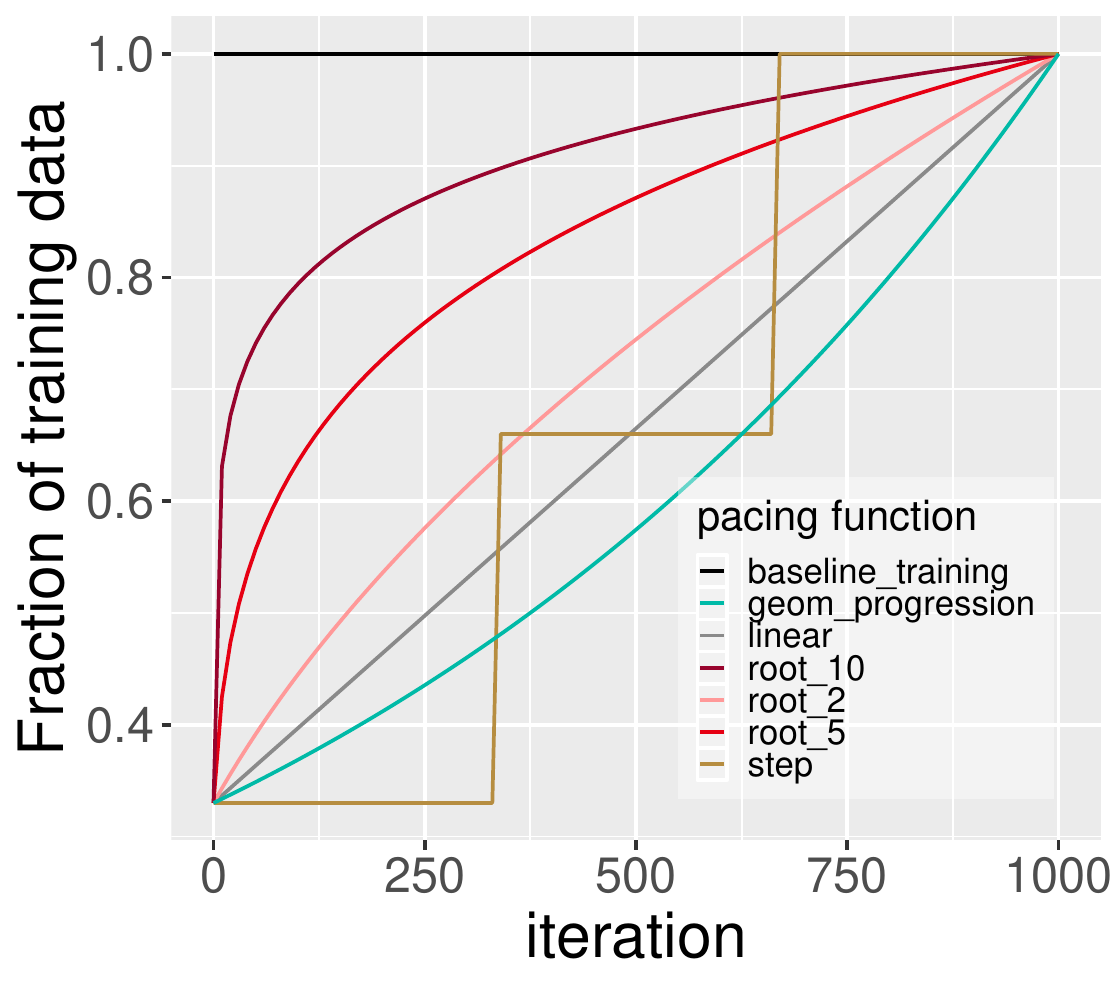}
}{\caption{Example with $\delta=0.33$ and $T=1000$.}\label{fig:pacing_functions}}

\end{floatrow}
\end{figure}

Assuming that we know the difficulty of each instance in our training set, we still need to define how are we going to transition from easy to hard instances. We use the concept of pacing functions $\pacing{}(s)$; they should each have the following properties~\cite{platanios2019competence,weinshall2018curriculum}: (i) start at an initial value of training instances $\pacing{}(0) = \delta$ with $\delta >0$, so that the model has a number of instances to train in the first iteration, (ii) be non-decreasing, so that harder instances are added to the training set, and, (iii) eventually all instances are available for sampling when it reaches $T$ iterations, $\pacing{}(T) = 1$.

As intuitively visible in the example in Figure~\ref{fig:pacing_functions}, we opted for pacing functions that introduce more difficult instances at different paces---while $root\_10$ introduces difficult instances very early (after 125 iterations, 80\% of all training data is available), $geom\_progression$ introduces them very late (80\% is available after $\sim 800$  iterations). We consider four different types of pacing functions, formally defined in Table~\ref{table:pacing_functions_definition}. The $step$ function~\cite{bengio2009curriculum,hacohen2019power,soviany2019image} divides the data into $S$ fixed sized groups, and after $\frac{T}{S}$ iterations a new group of instances is added, where $S$ is a hyperparameter. A more gradual transition was proposed by Platanios et. al~\cite{platanios2019competence}, by adding a percentage of the training dataset linearly with respect to the total of CL iterations $T$, and thus the slope of the function is $\frac{1-\delta}{T}$ ($linear$ function). They also proposed $root\_n$ functions motivated by the fact that difficult instances will be sampled less as the training data grows in size during training. By making the slope inversely proportional to the current training data size, the model has more time to assimilate difficult instances. Finally, we propose the use of a geometric progression that instead of quickly adding difficult examples, it gives easier instances more training time.
\section{Experimental Setup}
\subsubsection*{Datasets}
We consider two large-scale information-seeking conversation datasets (cf. Table \ref{table:dataset_stats}) that allow the training of neural ranking models for conversation response ranking. \msdialog{}\footnote{\msdialog{} is available at~\url{https://ciir.cs.umass.edu/downloads/msdialog/}}~\cite{qu2018analyzing} contain 246K context-response pairs, built from 35.5K information seeking conversations from the Microsoft Answer community, a question-answer forum for several Microsoft products. \mantis{}\footnote{\mantis{} is available at~\url{https://guzpenha.github.io/MANtIS/}}~\cite{penha2019introducing} was created by us and contains 1.3 million context-response pairs built from conversations of 14 different sites of \stack{}. Each \mantis{} conversation fulfills the following conditions: (i) it takes place between exactly two users (the information \emph{seeker} who starts the conversation and the information \emph{provider}); (ii) it consists of at least 2 utterances per user; (iii) one of the provider's utterances contains a hyperlink, providing grounding; (iv) if the final utterance belongs to the seeker, it contains positive feedback. We created \mantis{} to consider \emph{diverse} conversations from different domains besides technical ones. We include \msdialog{}~\cite{qu2018analyzing,yang2018response,InforSeek_User_Intent_Pred} here as a widely used benchmark.

\vspace{-0.5cm}
\begin{table}[]	
% \scriptsize
% \tiny
\small
\caption{Dataset used. $\set{U}$ is the dialogue context, $r$ a response and $u$ an utterance.}
\label{table:dataset_stats}
\centering
\begin{tabular}{p{6cm}rrrrrr}
\toprule
 & \multicolumn{3}{c}{\textbf{\msdialog{}}} & \multicolumn{3}{c}{\textbf{\mantis{}}} \\ 
 \midrule
Number of domains & \multicolumn{3}{c}{75} & \multicolumn{3}{c}{14} \\ 
\midrule
 & \multicolumn{1}{l}{Train} & \multicolumn{1}{l}{Valid} & \multicolumn{1}{l}{Test} & \multicolumn{1}{l}{Train} & \multicolumn{1}{l}{Valid} & \multicolumn{1}{l}{Test} \\ 
\midrule
Number of $(\set{U},r)$ pairs & 173k & 37k & 35k & 904k & 199k & 197k \\ 
Number of candidates per $\set{U}$ & 10 & 10 & 10 & 11 & 11 & 11 \\ 
Average number of turns & 5.0 & 4.8 & 4.4 & 4.0 & 4.1 & 4.1 \\ 
Average number of words per $u$ & 55.8 & 55.8 & 52.7 & 98.2 & 107.2 & 110.4 \\ 
Average number of words per $r$ & 67.3 & 68.8 & 67.7 & 91.0 & 100.1 & 94.6 \\ 
\bottomrule
\end{tabular}
\vspace{-6mm}%
\end{table}

\subsubsection*{Implementation Details}
As strong neural ranking model for our experiments, we employ BERT~\cite{devlin2019bert} for the conversational response ranking task. We follow recent research in IR that employed fine-tuned BERT for retrieval tasks~\cite{nogueira2019passage,yang2019simple} and obtain strong baseline (i.e., no CL) results for our task. The best model by Yang et. al~\cite{yang2018response}, which relies on external knowledge sources for \msdialog{}, achieves a MAP of 0.68 whereas our BERT baselines reaches a MAP of 0.71 (cf. Table~\ref{table:scoring_functions_result}). We fine-tune BERT\footnote{We use the PyTorch-Transformers implementation \url{https://github.com/huggingface/pytorch-transformers} and resort to \textit{bert-base-uncased} with default settings.} for sentence classification, using the CLS token\footnote{The BERT authors suggest CLS as a starting point for sentence classification tasks~\cite{devlin2019bert}.}; the input is the concatenation of the dialogue context and the candidate response separated by SEP tokens. When training BERT we employ a balanced number of relevant and non-relevant context and response pairs\footnote{We observed similar results to training with 1 to 10 ratio in initial experiments.}. We use cross entropy loss and the Adam optimizer~\cite{kingma2014adam} with learning rate of $5e-5$ and $\epsilon = 1e-8$. 

For \stdsm{}, as word embeddings we use pre-trained fastText\footnote{\url{https://fasttext.cc/docs/en/crawl-vectors.html}} embeddings with 300 dimensions and a maximum length of $k=20$ words of dialogue contexts and responses. For \stdbm{}, we use default values\footnote{\url{https://radimrehurek.com/gensim/summarization/bm25.html}} of $k_1=1.5$, $b=0.75$ and $\epsilon=0.25$. For CL, we fix $T$ as 90\% percent of the total training iterations---this means that we continue training for the final 10\% of iterations after introducing all samples---and the initial number of instances $\delta$ as 33\% of the data to avoid sampling the same instances several times.

\vspace{-0.5cm}
\subsubsection*{Evaluation} %%%%% CLAUDIA
To compare our strategies with the baseline where no CL is employed, for each approach we fine-tune BERT five times with different random seeds---to rule out that the results are observed only for certain random weight initialization values---and for each run we select the model with best observed effectiveness on the development set. The best model of each run is then applied to the test set. We report the effectiveness with respect to Mean Average Precision (MAP) like prior works~\cite{wu2017sequential,yang2018response}. We perform paired Student's t-tests between each scoring/pacing-function variant and the baseline run without CL.
\section{Results}
\begin{figure}[t]
% \captionsetup{margin=1cm}
\begin{floatrow}
\ffigbox[0.6\textwidth]{%
  \includegraphics[width=0.6\textwidth,left]{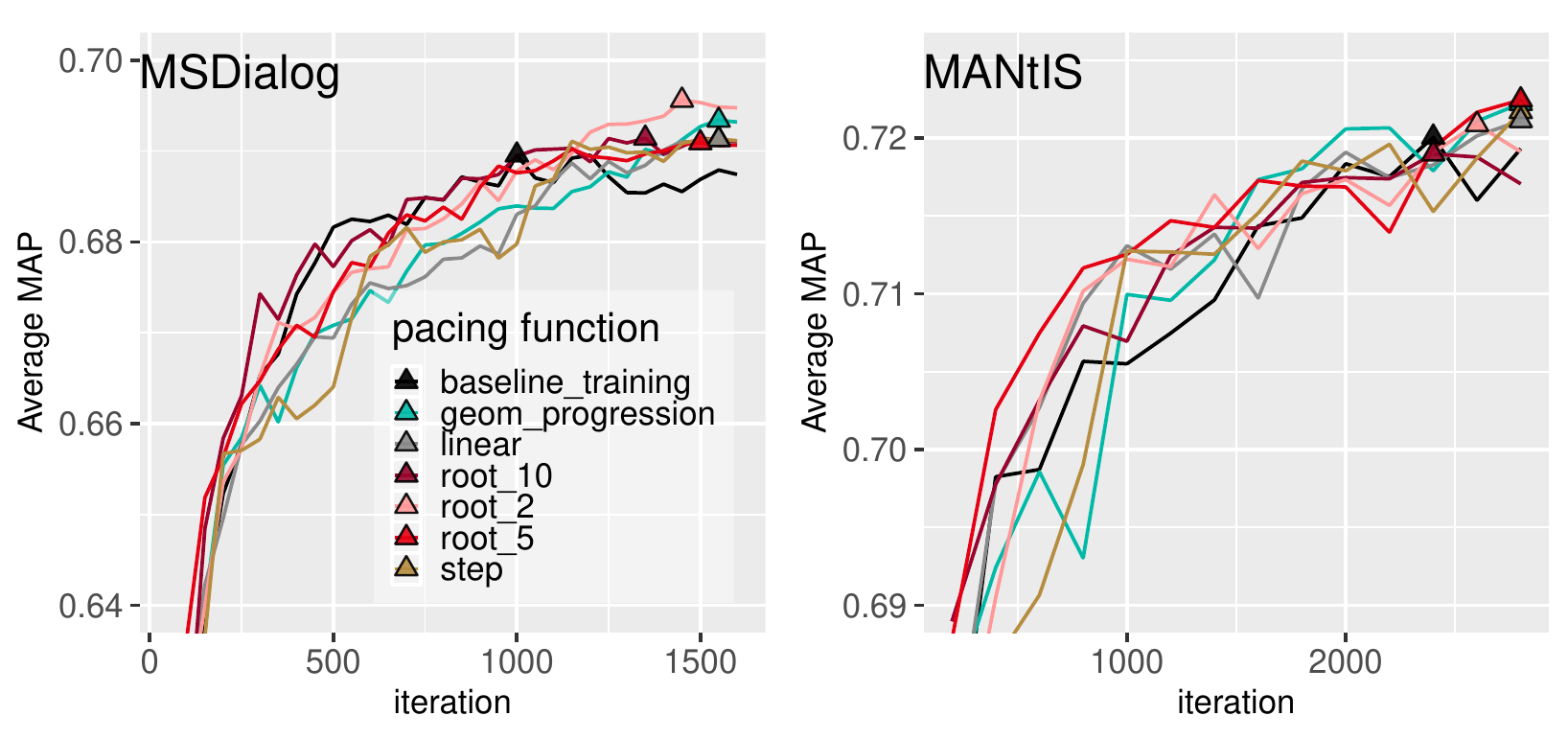}
}{\caption{Average development MAP for 5 different runs, using different curriculum learning pacing functions. $\bigtriangleup$ is the maximum observed MAP.}\label{fig:results_pacing}}

\ffigbox[0.39\textwidth]{%
  \includegraphics[width=0.39\textwidth,left]{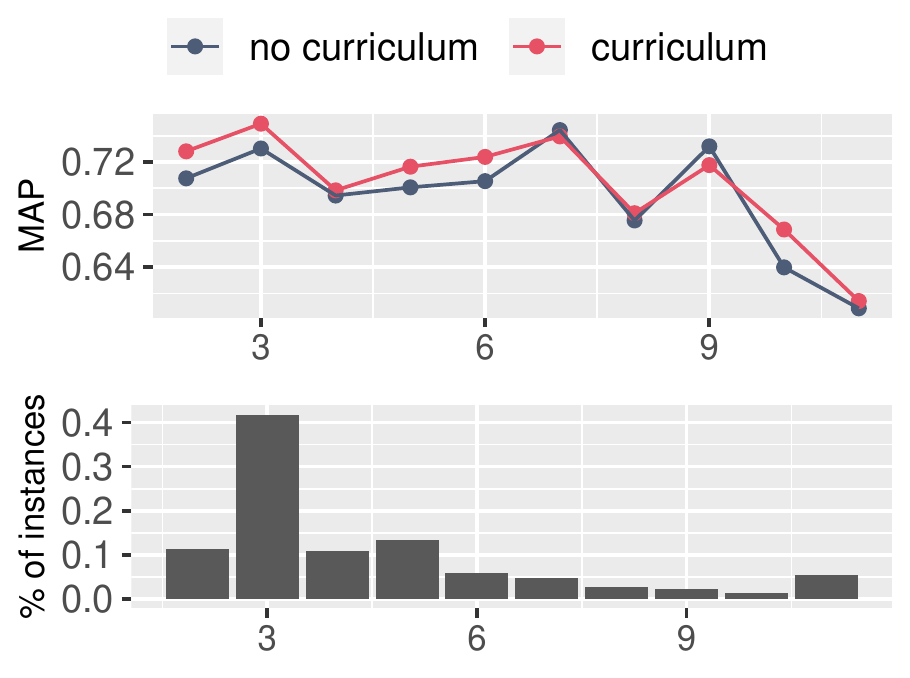}
}{\caption{\msdialog{} test set MAP of curriculum learning and baseline by number of turns.}\label{fig:error_analysis_turns}}
\end{floatrow}
\end{figure}

We first report the results for the pacing functions (Figure~\ref{fig:results_pacing}) followed by the main results (Table~\ref{table:scoring_functions_result}) comparing different scoring functions. We finish with an error analysis to understand when CL outperforms our no-curriculum baseline.

\subsubsection{Pacing Functions} In order to understand how CL results are impacted by the pace we go from easy to hard instances, we evaluate the different proposed \textit{pacing functions}. We display the evolution of the development set MAP (average of 5 runs) during training on Figure \ref{fig:results_pacing} (we use development MAP to track effectiveness during training). We fix the scoring function as \avgbertpred{}; this is the best performing scoring function, more details in the next section. We see that the pacing functions with the maximum observed average MAP are $root\_2$ and $root\_5$ for \msdialog{} and \mantis{} respectively\footnote{If we increase the $n$ of the root function to bigger values, e.g. $root\_10$, the results drop and get closer to not using CL. This is due to the fact that higher $n$ generate root functions with a similar shape to standard training, giving the same amount of time to easy and hard instances (cf. Figure~\ref{fig:pacing_functions}).}. The other pacing functions, \textit{linear}, \textit{geom\_progression} and \textit{step}, also outperform the standard training baseline with statistical significance on the test set and yield similar results to the \textit{root\_2} and \textit{root\_5} functions.

Our results are aligned with previous research on CL~\cite{platanios2019competence}, that giving more time for the model to assimilate harder instances (by using a root pacing function) is beneficial to the curriculum strategy and is better than no CL with statistical significance on both development and test sets. For the rest of our experiments we fix the pacing function as $root\_2$, the best pacing function for \msdialog{}. Let's now turn to the impact of the scoring functions.

\subsubsection{Scoring Functions}

\begin{table}[ht]
% \scriptsize
% \tiny
\centering
\caption{Test set MAP results of 5 runs using different curriculum learning scoring functions. Superscripts $^{\dagger}/^{\ddagger}$ denote statistically significant improvements over the baseline where no curriculum learning is applied ($\scoring{}=random$) at 95\%/99\% confidence intervals. Bold indicates the highest MAP for each line.}
\label{table:scoring_functions_result}
\begin{tabular}{@{}p{0.9cm}llllllll@{}}
\toprule
\multicolumn{9}{c}{\msdialog{}} \\ \midrule
% \backslashbox{r}{}
run 
& $random$ & \numberutterances & \avgnumwordsq & \avgnumwordsd & \stdsm & \stdbm & \avgbertpred & \avgbertloss \\ \midrule
1 & 0.7142 & 0.7220 $^{\dagger}$ & 0.7229 $^{\dagger}$ & 0.7182 & 0.7239 $^{\dagger \ddagger}$ & 0.7175 & \textbf{0.7272} $^{\dagger \ddagger}$ & 0.7244 $^{\dagger \ddagger}$ \\
2 & 0.7044 & 0.7060 & 0.7053 & 0.6968 & 0.7032 & 0.7003 & \textbf{0.7159 }$^{\dagger \ddagger}$ & 0.7194 $^{\dagger \ddagger}$ \\
3 & 0.7126 & 0.7215 $^{\dagger}$ & 0.7163 & 0.7171 & 0.7174 & 0.7159 & \textbf{0.7296} $^{\dagger \ddagger}$ & 0.7225 $^{\dagger \ddagger}$ \\
4 & 0.7031 & 0.7065 & 0.7043 & 0.6993 & 0.7026 & 0.6949 & 0.7154 $^{\dagger \ddagger}$ & \textbf{0.7204} $^{\dagger \ddagger}$ \\
5 & 0.7148 & 0.7225 $^{\dagger}$ & 0.7203 & 0.7169 & 0.7171 & 0.7134 & 0.7322 $^{\dagger \ddagger}$ & \textbf{0.7331} $^{\dagger \ddagger}$ \\ \midrule
AVG & 0.7098 & 0.7157 & 0.7138 & 0.7097 & 0.7128 & 0.7084 & \textbf{0.7241} & 0.7240 \\ \midrule
SD & 0.0056 & 0.0086 & 0.0086 & 0.0106 & 0.0095 & 0.0101 & 0.0079 & 0.0055 \\ \midrule
\multicolumn{9}{c}{\mantis{}} \\ \midrule
1 & 0.7203 & 0.7192 & 0.7198 & 0.7194 & 0.7166 & 0.7200 & 0.7257 $^{\dagger \ddagger}$ &\textbf{0.7268} $^{\dagger \ddagger}$ \\
2 & 0.6984 & 0.6993 & 0.6989 & 0.6996 & 0.6964 & 0.7009 & \textbf{0.7067} $^{\dagger \ddagger}$ & 0.7051 $^{\dagger \ddagger}$ \\
3 & 0.7200 & 0.7197 & 0.7134 & 0.7206 & 0.7153 & 0.7153 & \textbf{0.7282} $^{\dagger \ddagger}$ & 0.7221 \\
4 & 0.7114 & 0.7117 & 0.7002 & 0.6978 & 0.7140 & 0.7084 & \textbf{0.7240} $^{\dagger \ddagger}$ & 0.7184 $^{\dagger \ddagger}$ \\
5 & 0.7156 & 0.7174 & 0.7193 $^{\dagger}$ & 0.7162 & 0.7147 & 0.7185 & \textbf{0.7264} $^{\dagger \ddagger}$ & 0.7258 $^{\dagger \ddagger}$ \\ \midrule
AVG & 0.7131 & 0.7135 & 0.7103 & 0.7107 & 0.7114 & 0.7126 & \textbf{0.7222} & 0.7196 \\ \midrule
SD & 0.0090 & 0.0085 & 0.0102 & 0.0111 & 0.0084 & 0.0079 & 0.0088 & 0.0088 \\ \bottomrule
\end{tabular}
\vspace{-0.5cm}
\end{table}
\vspace{-0.5cm}

The most critical challenge of CL is defining a measure of difficulty of instances. In order to evaluate the effectiveness of our scoring functions we report the test set results across both datasets in Table~\ref{table:scoring_functions_result}. We observe that the scoring functions which do not use the relevance labels $\set{Y}$ are not able to outperform the no CL baseline (\textit{random} scoring function). They are based on features of the dialogue context $\set{U}$ and responses $\set{R}$ that we hypothesized make them difficult for a model to learn. Differently, for \avgbertloss{} and \avgbertpred{} we observe statistically significant results on both datasets across different runs. They differ in two ways from the unsuccessful scoring functions: they have access to the training labels $\set{Y}$ and the difficulty of an instance is based on what a previously trained model determines to be hard, and thus not our intuition.

Our results bear resemblance to Born Again Networks~\cite{furlanello2018born}, where a student model which is identical in parameters and architecture to the teacher model outperforms the teacher when trained with knowledge distillation~\cite{hinton2015distilling}, i.e., using the predictions of the teacher model as labels for the student model. The difference here is that instead of transferring the knowledge from the teacher to the student through the labels, we transfer the knowledge by imposing a structure/order on the training set, i.e. curriculum learning.

\subsubsection{Error Analysis}
\begin{figure}[t]
    \centering
    \includegraphics[width=1\textwidth]{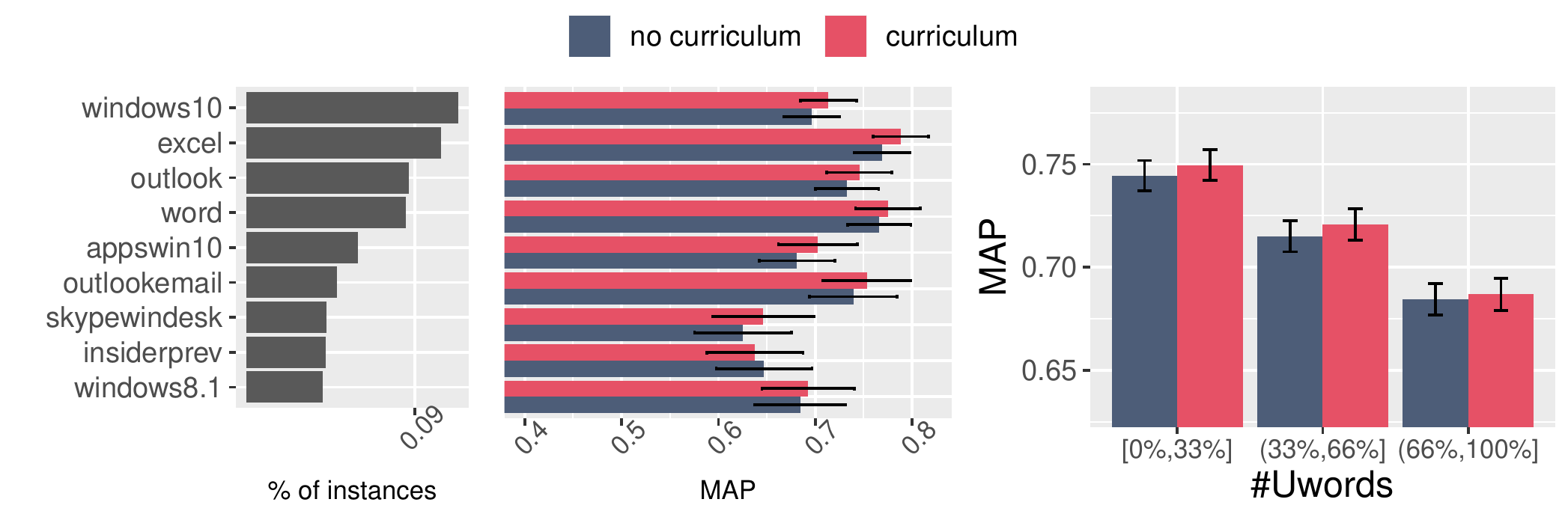}
    \caption{Test set MAP for \msdialog{} across different domains (left) and instances' difficulty (right)  according to \avgnumwordsd{} for curriculum learning and the baseline.}
    \label{fig:error_analysis}
\end{figure}

In order to understand when CL performs better than random training samples, we fix the scoring (\avgbertpred{}) ad pacing function (\textit{root\_2}) and explore the test set effectiveness along several dimensions (cf. Figures~\ref{fig:error_analysis_turns} and~\ref{fig:error_analysis}). We report the results only for \msdialog{}, but the trends hold for \mantis{} as well.

We first consider the number of turns in the conversation in Figure~\ref{fig:error_analysis_turns}. CL outperforms the baseline approach for the types of conversations appearing most frequently (2-5 turns in \msdialog{}). The CL-based and baseline effectiveness drops for conversations with a large number of turns. This can be attributed to two factors: (1) employing pre-trained BERT in practice allows only a certain maximum number of tokens as input, so longer conversations can lose important information due to truncating; (2) for longer conversations it is harder to identify the important information to match in the history, i.e information spread.

Next, we look at different conversation domains in Figure~\ref{fig:error_analysis} (left), such as \textit{physics} and \textit{askubuntu}---are the gains in effectiveness limited to particular domains? The error bars indicate the confidence intervals with confidence level of 95\%. We list only the most common domains in the test set. The gains of CL are spread over different domains as opposed to concentrated on a single domain.

Lastly, using our scoring functions we sort the test instances and divide them into three buckets: first 33\% instances, 33\%--66\% and 66\%--100\%. In Figure~\ref{fig:error_analysis} (right), we see the effectiveness of CL against the baseline for each bucket using \avgnumwordsq{} (the same trend holds for the other scoring functions). As we expect, the bucket with the most difficult instances according to the scoring function is the one with lowest MAP values. Finally, the improvements of CL over the baseline are again spread across the buckets, showing that CL is able to improve over the baseline for different levels of difficulty.

% ~\footnote{We take a further look into why this does not hold for \avgnumwordsd in \mantis{} on the Appendix.}. 

\section{Conclusions}
\vspace{-0.2cm}
In this work we studied whether CL strategies are beneficial for neural ranking models. We find supporting evidence for curriculum learning in IR. Simply reordering the instances in the training set using a difficulty criteria leads to effectiveness improvements, requiring no changes to the model architecture---a similar relative improvement in MAP has justified novel neural architectures in the past~\cite{wu2017sequential,zhang2018modeling,zhou2018multi,tao2019one}. Our experimental results on two conversation response ranking datasets reveal (as one might expect) that it is best to use all available information $(\set{U},\set{R},\set{Y})$ as evidence for instance difficulty. Future work directions include considering other retrieval tasks, different neural architectures and an investigation of the underlying reasons for CL's workings.

\small
\subsubsection*{Acknowledgements}
This research has been supported by NWO projects SearchX (639.022.722) and NWO Aspasia (015.013.027).
\bibliographystyle{splncs04}
\bibliography{references.bib}

\end{document}